\documentclass{aa}
\usepackage{graphics,epsfig,amsmath,amssymb,amstext}

\def\d {\mathrm{d}}
\def\Msun{\hbox{$\mathrm{M}_{\odot}$}}
\def\libeb{$^{6}$LiBeB}

\begin{document}

\thesaurus{12(02.01.1; 02.14.1; 09.03.2; 10.01.1)}

\title{Superbubbles as the source of $^{6}$Li, Be and B in the early Galaxy}

\author{E. Parizot \and L. Drury}

%\offprints{E.~Parizot}

\institute{Dublin Institute for Advanced Studies, 5 Merrion Square, 
Dublin 2, Ireland\\ e-mail: parizot@cp.dias.ie; ld@cp.dias.ie}

\date{Received date: 1 April 1999;accepted date: 31 May 1999}

\authorrunning{E. Parizot \& L. Drury}
\titlerunning{Superbubbles as sources of $^{6}$LiBeB}
\maketitle

\begin{abstract}

We investigate the spallative production of the light elements, Li, Be 
and B (LiBeB), associated with the evolution of a superbubble (SB) 
blown by repeated supernovae (SNe) in an OB association.  It is shown 
that if about ten percent of the SN energy can power the acceleration 
of particles from the material inside the SB, the observed abundances 
of LiBeB in halo stars, as a function of O, can be explained in a 
fully consistent way over several decades of metallicity.  In this 
model, the energetic particles (EPs) reflect the SB material, which is 
a mixing of the ejecta of previous SNe and of the swept-up ISM gas 
evaporated off the shell.  We investigated two different energy 
spectra for the EPs: the standard cosmic ray source spectrum, or `SNR 
spectrum', and a specific `SB spectrum', $E^{-\alpha}\exp(-E/E_{0})$, 
where $\alpha = 1.$--1.5 and $E_{0}$ is of order a few hundreds of 
MeV/n, as results from the SB acceleration mechanism of Bykov \& 
Fleishman (1992).  While the latter spectrum is more efficient in 
producing LiBeB, the SNR spectrum can be reconciled with the 
observational data if an imperfect mixing of the SN ejecta with the 
rest of the SB material and/or a selective acceleration is invoked 
(enhancing the C and O abundance amongst the EPs by a factor of $\sim 
6$).  One of the main consequences of our model is that the observed 
linear growth of Be and B abundances as a function of Fe/H expresses a 
dilution line rather than a continuous, monotonic increase of the 
metallicity.  We propose an observational test of this feature.  We 
also show that the recent $^{6}$Li observations in halo stars fit 
equally well in the framework of the SB model.  Finally, we conjecture 
the existence of two sets of low-metallicity stars, differing in their 
Be/Fe or B/O abundance ratios, resulting from a `bimodal' LiBeB 
production in the Galaxy, namely from correlated (in SBs) or isolated 
SN explosions.


\end{abstract}

\section{Introduction}

It is generally accepted that all the elements in the universe have 
been produced (synthesized) in stars or stellar explosions, except the 
so-called primordial (H, He and part of the $^{7}$Li), and light 
elements (Li, Be and B).  The former class of nuclei is satisfactorily 
accounted for by Big Bang nucleosynthesis models, while the latter is 
thought to be produced by non thermal processes in the interstellar 
medium (ISM).  Indeed, $^{6}$Li, Be and B nuclei (\libeb) are actually 
destroyed rather than produced through the thermal nuclear processes 
in stellar interiors.  The first, and most natural thesis concerning 
the origin of these light elements (Reeves et al.  1970, Meneguzzi et 
al.  1971) claimed that they were produced through spallation 
reactions of C and O nuclei induced by the interaction of the Galactic 
cosmic rays (mostly composed of protons and alpha particles, 
especially in the early, metal-poor Galaxy) with the ambient ISM. This 
process is referred to as the Galactic cosmic ray nucleosynthesis 
(GCRN).  As the Galaxy evolves, the ISM gets richer and richer in C 
and O, making the spallative nucleosynthesis more and more efficient 
(one particular energetic proton has more chance to hit a C or O 
nucleus).  As a consequence of the production \emph{rates} being 
proportional to the ambient C and O abundance, the GCRN models predict 
an increase of the \libeb~abundance in the ISM, and thus in stars, 
proportional to the square of the metallicity, defined either as the O 
or Fe abundance.

However, the measurement of Be, B, and most recently $^{6}$Li 
abundances in very metal-poor stars of the Galactic halo has 
considerably changed this picture (see e.g.  Vangioni-Flam et al., 
1998, for a review).  Two important problems have arisen.  The first 
one is qualitative: the increase of Be and B is found to be 
proportional to the Fe abundance (instead of its square) for stars of 
metallicity ranging from $10^{-3}$ to $10^{-1}$ times the solar 
metallicity.  The second problem is quantitative: the amount of Be and 
B nuclei effectively observed in halo stars requires an extremely 
large number of spallation reactions, as compared to what can be 
expected from the standard GCRN scenario.  This, in turn, implies that 
an enormous amount of energy has been imparted to the cosmic rays in 
the early Galaxy.

The most natural answer to both of these problems (qualitative and 
quantitative) is to assume that the energetic particles (EPs) inducing 
the spallation reactions are not representative of the ambient ISM at 
the time of their acceleration, as in the case of GCRN, but rather 
show a much richer abundance in C and O, so that the 
\libeb~nucleosynthesis is mainly due to energetic C and O nuclei being 
spalled in flight on H and He nuclei at rest in the ISM (so-called 
\emph{inverse spallation}), rather than the opposite, i.e.  energetic 
H and He interacting with C and O nuclei at rest (so-called 
\emph{direct spallation}).  In this way, the \libeb~spallative 
production rate in the Galaxy depends only on the power, composition 
and energy spectrum of the EPs, not on the composition of the ambient 
ISM. It is therefore directly proportional to the EP acceleration rate 
in the Galaxy, and thus presumably to the supernova explosion rate.  
As a consequence, the abundance of \libeb~increases proportionally to 
that of C and O, in agreement with the observations.  Moreover, since 
the EPs are C and O rich, the efficiency of the nucleosynthetical 
process is much higher than in the case of GCRN, where a lot of energy 
is imparted to (and lost in) protons and alpha particles which never 
hit any C or O nucleus in the ISM to produce light elements.

This idea of a process dominated by inverse spallation reactions has 
been discussed in a number of phenomenological papers (e.g.  Cass\'e 
et al.  1995, Ramaty et al.  1996, 1997, Vangioni-Flam et al.  1998).  
However, while this solution framework clearly improves the situation, 
some important questions still need to be answered: what is the 
composition of the EPs responsible for the production of the 
\libeb~nuclei observed in halo stars?  What is the EP energy 
spectrum?  What is their source of energy?  What is the acceleration 
mechanism?  How efficient is it?  Where does the acceleration take 
place?  Where do the spallation reactions take place?  All these 
questions are linked together in more or less complex a way, as the 
energy requirement of the \libeb~production depends on the composition 
and spectrum of the EPs, both of which depend in turn on the 
acceleration site and mechanism.

In a pair of recent papers (Parizot \& Drury 1999a,b), we have 
investigated in detail two light elements production mechanisms 
associated with the explosion of a supernova in the ISM, and 
calculated the total amount of Be they produce.  While both of these 
processes are \emph{primary} and therefore reproduce the observed 
linear growth of \libeb~abundances as a function of metallicity, we 
showed that they fail to solve the quantitative part of the problem, 
and must therefore be abandoned.  Analyzing the reasons for this 
failure, we pointed out one possible alternative scenario, based on 
the acceleration of enriched material within superbubbles.  Such a 
model, focusing on the collective effect of many supernovae occurring 
repeatedly in OB associations, rather than on individual supernovae, 
is now investigated in detail in this paper.

\section{Description of the model}
\label{Model}

\subsection{Overview}
\label{Overview}

We are interested in the total production of Li, Be and B associated 
with the evolution of a typical superbubble (SB) generated by the 
supernova activity of an OB association in the interstellar medium.  
The basic idea is the following: when several SN explosions occur 
locally both in space and time, a superbubble forms and grows, 
surrounded by a shell of swept-up ISM gas.  Its interior is filled 
with hot ($T > 10^{6}$~K), tenuous ($n\la 10^{-2}$) gas, made of i) 
the ejecta of previous supernovae and possible winds of massive stars 
and ii) the interstellar material evaporated off the shell or embedded 
clouds.  When a new supernova explodes within an already formed 
superbubble, part of its energy powers some acceleration process 
(usual SN shock acceleration or any other specific process) and goes 
into EPs which then induce spallation reactions by which a certain 
amount of LiBeB is produced.  The specificity of the SB model is 
twofold: i) since the interior of the SB receives the ejecta of 
previous SNe, the composition of the EPs is naturally enriched in C 
and O, even in the absence of selective acceleration, which makes the 
LiBeB production particularly efficient and ii) the energy spectrum of 
the EPs may be different from the usual CR source spectrum, as a 
result of a possibly different acceleration mechanism due to the 
specific physical conditions prevailing inside the SB (Bykov \& 
Fleishman, 1992; Bykov, 1995,1999).

An other specific feature of our SB model is that it implies a 
discontinuous increase of the LiBeB abundance in the ISM. Unlike the 
`ISM models' where both metals and light elements build up 
continuously as the stars process the primordial Galactic gas in the 
ISM and progressively increase its metallicity, the SB model provides 
a locally strong increase of the metallicity and the LiBeB abundance, 
resulting from the accumulation of the ejecta of many massive stars in 
a small region of space (the SB), on a relatively short time scale (a 
few tens of Myr).  Anticipating the main result of this paper, we 
emphasize that a typical SB developing in the very early Galaxy (with 
a mean ISM metallicity $Z = 10^{-4}Z_{\odot}$) provides after, say, 
30~Myr, about $8\,10^{4}\,\Msun$ of material with metallicity $Z = 
10^{-1}Z_{\odot}$ \emph{and} the correct (i.e.  observed) LiBeB 
abundances relative to this metallicity.  When the SB then breaks up, 
these $\sim 8\,10^{4}\,\Msun$ of high metallicity material mix with 
the ambient ISM, whose metallicity is three orders of magnitude lower 
in our canonical example.  Depending on the mixing or dilution 
efficiency, the SB gives rise to concentrations of matter with various 
metallicities, extending from $Z = 10^{-1}Z_{\odot}$ (virtually no 
dilution) to $Z = 10^{-4}Z_{\odot}$ (dilution of the SB material by a 
large amount of ISM gas).  The next generation of stars formed from 
the collapse of this variously enriched gas will thus show very 
different metallicities, but always the same abundance ratios (such as 
Be/O, or B/Fe), namely those of the parent SB itself.  This model 
therefore predicts constant abundance ratios (including those 
involving \libeb) over several decades of metallicity, in conformity 
with the observations.  But the linear increase of, say, the Be 
abundance as a function of Fe/H, now has a very different origin: it 
is essentially a \emph{dilution line}, not a `constant growth line', 
or `accumulation line'.

In addition, this also implies that the relation between the age and 
metallicity of the metal-poor stars is no longer monotonic and 
one-to-one.  Two stars formed at the same time from the gas processed 
in a given SB can indeed show very different metallicities, depending 
on the dilution of this gas with the ambient ISM before the 
gravitational collapse.  This is in contrast with the Galactic 
chemical evolution models where a steady, continuous enrichment of the 
interstellar gas is assumed.  An observational test of this important 
feature is suggested in Sect.~\ref{Conclusion}.

\subsection{The superbubble model}

The detailed structure and evolution law of a given superbubble 
obviously depends on the local density profile as well as on the 
history of the SN explosions, determined by the particular stellar 
content and stellar formation history of the OB association.  However, 
since we are only interested in the mean LiBeB yield per supernova or 
per superbubble, we shall make some general simplifying assumptions to 
calculate the output to be expected from a `typical superbubble'.

Following previous work on the evolution of superbubbles in the ISM 
(e.g.  Mac Low \& McCray 1988), we shall assume a continuous power 
supply from SN explosions regularly spread in time in a localized OB 
association, within a homogeneous ISM. The resulting SB is therefore 
spherically symmetric and can be treated by the standard self-similar 
model of Weaver et al.  (1977) for a pressure-driven wind bubble.  The 
expressions for the SB radius, mass, internal mean density and 
temperature then read:

\begin{equation}
	R_{\mathrm{SB}}(t) =
	(66~\mathrm{pc})\,L_{38}^{1/5}n_{0}^{-1/5}t_{\mathrm{Myr}}^{3/5}
	\label{RSB}
\end{equation}
\begin{equation}
	M_{\mathrm{SB}}(t) = 
	(1600~\Msun)\,L_{38}^{27/35}n_{0}^{-2/35}t_{\mathrm{Myr}}^{41/35}
	\kappa_{0}^{2/7}
	\label{MSB}
\end{equation}
\begin{equation}
	n_{SB}(t) = (1.6\,10^{-2}~\mathrm{cm}^{-3})\,
	L_{38}^{6/35}n_{0}^{19/35}t_{\mathrm{Myr}}^{-22/35}\kappa_{0}^{2/7}
	\label{nSB}
\end{equation}
\begin{equation}
	T_{\mathrm{SB}}(t) = (5.3\,10^{6}~\mathrm{K})\,
	L_{38}^{8/35}n_{0}^{2/35}t_{\mathrm{Myr}}^{-6/35}\kappa_{0}^{-2/7}
	\label{TSB}
\end{equation}
where $L_{38}$ is the mechanical luminosity of the OB association 
(assumed constant) in units of $10^{38}$~erg~s$^{-1}$, $n_{0}$ is the 
ISM number density in cm$^{-3}$, and $t_{\mathrm{Myr}}$ is the SB age 
in Myr.  The mass of the SB is mainly due to conductive evaporation 
from the shell.  Following Shull \& Saken (1995), we have multiplied 
the classical conductivity by a dimensionless scaling factor 
$\kappa_{0} \le 1$ to account for possible magnetic suppression.  As 
can be seen in the above equations, the dependence of 
$M_{\mathrm{SB}}$ and $n_{\mathrm{SB}}$ on $\kappa_{0}$ is rather 
weak, so that the magnetic suppression would have to be very strong 
($\kappa_{0}\ll 1$) to produce a large diminution of the SB mass and 
density, at a given time.

Higdon et al.  (1998) have distinguished between magnetic and 
non-magnetic models in a different way: referring to Tomisaka (1992)'s 
numerical model of magnetized SBs, they assumed that the mass injected 
into the SB averages about 45~\Msun~per supernova.  This implies a 
much lower SB mass, so that the enriched ejecta of the SNe are less 
diluted and the composition of the SB material to be accelerated is 
correspondingly much richer in C and O. However, this model does not 
seem very realistic, considering the extremely low SB density which it 
implies~: $n_{\mathrm{SB}}\sim 2\,10^{-4}~\mathrm{cm}^{-3}$ for 
$t_{\mathrm{SB}} = 50$~Myr (Higdon et al., 1998), to be compared with 
typical values inferred from the observations of order $\ga 
10^{-2}~\mathrm{cm}^{-3}$ (e.g.  Brown et al., 1995; Bomans et al., 
1997; Walter et al., 1998), in better agreement with Eq.~(\ref{nSB}).  
Moreover, the value of 45~\Msun used by Tomisaka is, on his own 
admission, arbitrary, and taken essentially to ensure the numerical 
stability of his code.  Therefore, we shall not use this low-mass 
model here, and hold on to the `canonical' SB model described by 
Eqs.~(\ref{RSB})--(\ref{nSB}).

The typical parameters which we use are the following~: $n_{0} = 
1\,\mathrm{cm}^{-3}$, $\kappa_{0} = 1$, and $L_{38} = 1$, which 
corresponds to one SN explosion every $\sim 3\,10^{5}$~yr (for an 
explosion energy $E_{\mathrm{SN}} = 10^{51}$~erg).  Assuming a 
lifetime of 30~Myr for the SB thus amounts to saying that the OB 
association provides a total of 100~SNe.  However, we also investigate 
less active SBs, with mechanical luminosities $L_{38} = 0.1$, and 
poorer OB associations with only a few tens of SNe.

\subsection{The production of light elements in superbubbles}
\label{LiBeBInSBs}

To calculate the LiBeB production associated with the SB evolution, we 
apply the time dependent model described in Parizot (1999).  To do so, 
we first determine the so-called injection function, $Q(E,t)$, which 
gives the power per unit energy imparted to the EPs, as well as their 
energy spectrum and composition as a function of time (for more 
details, see Parizot \& Drury, 1999b, where the model is applied in 
much the same way to the spallative nucleosynthesis in supernova 
remnants).  Then we let the EPs interact with the surrounding medium, 
in which they suffer energy losses and nuclear reactions, some of 
which lead to LiBeB production.  We thus deduce the production rates 
as a function of time, and integrate them to obtain the total yields 
in the light elements: $\mathcal{N}_{\mathrm{Li}}$, 
$\mathcal{N}_{\mathrm{Be}}$ and $\mathcal{N}_{\mathrm{B}}$.

We assume that the power imparted to the EPs is, at any time, a 
fraction $\theta$ of the total mechanical power supplied by the OB 
association.  In the early Galaxy, the contribution of the winds of 
massive stars is thought to be very small, because of the low stellar 
metallicity, so the available energy is mainly due to the SN 
explosions.  In conformity with standard shock acceleration models, we 
assume $\theta = 0.1$, but the results can be scaled straightforwardly 
to any other value.  As for the EP energy spectrum, we investigated 
two different forms: i) a standard shock acceleration spectrum, i.e.  
$Q(p) \propto p^{-4}$, thought to be also the CR source spectrum, and 
called here the `SNR spectrum' to emphasize that it originates at an 
isolated supernova remnant, and ii) a `SB spectrum', $Q(E) \propto 
E^{-\alpha}\exp(-E/E_{0})$, where $\alpha = 1$ or 1.5, and $E_{0}$ is 
a cut-off energy of typically a few hundreds of MeV/n.  This spectrum 
is taken as an approximate of the time-dependent source spectrum 
derived from an acceleration model relevant to the specific physical 
conditions prevailing within a SB (Bykov \& Fleishman, 1992; Bykov, 
1995,1999).

While the SB evolves, more and more particles get accelerated and 
contribute to the LiBeB production rates.  We assume that the EPs are 
confined within the SB during its whole lifetime, as a consequence of 
the magnetic waves and strong magnetic turbulence.  This implies that 
they interact with the enriched SB material.  However, this assumption 
is of little influence on the total LiBeB yields, because it is found 
that the spallative nucleosynthesis is dominated by inverse spallation 
processes, which are essentially independent of the target 
composition.  Moreover, since the SB density is very low, most of the 
EPs have not yet been spalled and still have a supernuclear energy at 
the end of the SB lifetime ($t = \tau_{\mathrm{SB}}$).  They then 
diffuse away in the surrounding medium, interact with the ISM gas in 
the dense shell or stay confined for a while within the dislocating 
SB. Again, their genuine fate is not crucial to our calculations, 
since the total LiBeB production (i.e.  integrated over time) is 
independent of both the target composition (as long as it is much 
poorer in C and O than the EPs are) and density.  For simplicity, we 
assume here that the target composition after $\tau_{\mathrm{SB}}$ is 
the ISM composition, and its density is 0.1~cm$^{-3}$.  This rather 
low value weakens the discontinuity of the production rates (see 
Fig.~\ref{Be-detail}) and ensures a smooth numerical behavior.  
However, when we want to show some specific behaviors such as the 
dependence of the production rates on the ambient density or the EP 
spectrum (Figs.~\ref{Be(n0)} and~\ref{Be(E0)}), we assume that the 
target density after $\tau_{\mathrm{SB}}$ is equal to 
$n_{\mathrm{SB}}(\tau_{\mathrm{SB}})$, which provides continuous 
curves.  On the other hand, it implies artificially long timescales, 
even longer than the age of the Galaxy.  In reality, once the SB 
dislocates and/or the EPs leave the SB interior, the mean density of 
the medium in which they propagate increases rapidly, making the 
spallation timescales much shorter (by the same factor as the density 
ratio).  Most probably, the main target for the EPs is the shell 
surrounding the SB, whose density is higher than the ambient ISM.

\subsection{The EP composition}

Most important also is the composition of the EPs.  Whatever the 
acceleration mechanism at work within the SB, the particles to be 
accelerated are those who stand in the hot bubble where the energy is 
released.  If the acceleration mechanism is not selective (i.e.  the 
acceleration efficiency is independent of the nuclear species), the EP 
composition is thus nothing but the composition of the material inside 
the SB, which is a mixing of the SN ejecta accumulated at time $t$ and 
the ISM material evaporated from the shell.  Higdon et al.  (1998) 
have argued that SN explosions occur mainly in the core of 
superbubbles where the metallicity is dominated by the SN ejecta.  
However, this does not mean that there is no dilution at all.  In the 
very early Galaxy, the ISM contains no metals, so the metallicity of 
the SB interior is certainly dominated by the SN ejecta, wherever the 
new explosion occurs and the particle acceleration takes place.  
However, the composition of the EPs still depends on the mixing of the 
ejecta with the ISM material evaporated from the shell.  Moreover, 
some metal-poor material may be evaporated from high density clouds 
embedded in the SB (e.g.  Cioffi \& Shull, 1991; Zanin \& 
Weinberger, 1997), close enough to the core to be accelerated directly 
by the SN shock.

To evaluate the degree of gas mixing, we assume that it occurs on a 
timescale $\tau_{\mathrm{mix}} = R_{\mathrm{SB}}/V_{\mathrm{mix}}$, 
where $V_{\mathrm{mix}}$ is a `turbulent velocity' of the order of the 
sound speed, $c_{\mathrm{s}}$, and compare $\tau_{\mathrm{mix}}$ with 
the age of the SB, $t$.  Calculating $c_{\mathrm{s}}$ from the SB 
temperature, Eq.~(\ref{TSB}), we obtain:
\begin{equation}
	\tau_{\mathrm{mix}} \simeq (0.67~\mathrm{Myr})\,
	\left(\frac{L_{38}\,t_{\mathrm{Myr}}}{n_{0}}\right)^{1/5}.
	\label{tauMix}
\end{equation}
The condition for the mixing time to be smaller than the SB age, 
i.e. $\tau_{\mathrm{mix}}<t$, then reads:
\begin{equation}
	t > (5\,10^{5}\,\mathrm{yr})\sqrt{\frac{L_{38}}{n_{0}}}.
	\label{MixingCondition}
\end{equation}

We thus find that the gas evaporated from the SB should have had 
enough time to mix with the SN ejecta as soon as $t\ga 5\,10^{5}$~yr, 
i.e.  after the explosion of the first or second SN. To take this 
result into account, we assume that the EP composition at any time 
reflects that of the accumulated SN ejecta diluted by and perfectly 
well mixed with the ISM gas evaporated at that time.  Note that this 
is a conservative assumption, as imperfect mixing and/or selective 
acceleration could raise the C and O abundance amongst the EPs, and 
thus increase the LiBeB yields.

%%%%%%%%%%%%%%
\begin{figure}
\resizebox{\hsize}{!}{\includegraphics{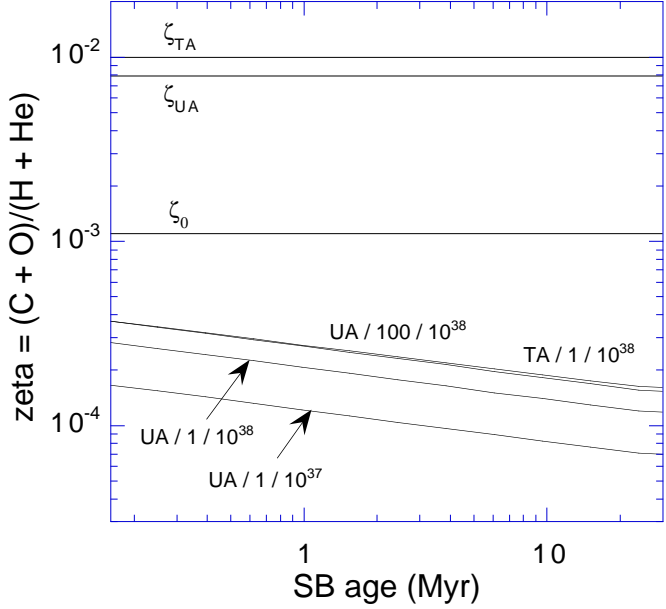}}
\caption{Evolution of the reduced metallicity, $\zeta$, over the 
superbubble lifetime (here, 30~Myr).  The labels indicate, 
respectively, the SN model used (from WW95), the ambient density (in 
cm$^{-3}$) and the mechanical power (in $\mathrm{erg~s}^{-1}$).  Also 
shown are the solar reduced metallicity ($\zeta_{0}$), and the reduced 
metallicity corresponding to the pure ejecta for models UA and TA of 
WW95 (the compositions have been averaged over an IMF of index 2.35).}
\label{zeta}
\end{figure}
%%%%%%%%%%%%%%

Figure~\ref{zeta} shows the evolution of the EP composition as a 
function of time, for different SN models, ambient densities and SB 
mechanical powers.  We use the explosion models of Woosley \& Weaver 
(1995, hereafter WW95) for OB stars of initial metallicity $Z = 
10^{-4}Z_{\odot}$ (models U), and $Z = 10^{-2}Z_{\odot}$ (models T), 
averaging all the yields over an initial mass function (IMF) of index 
$x$.  We found that the results do not change significantly for any 
reasonable value of $x$, so we only show the results obtained with the 
Salpeter index $x = 2.35$.  The efficiency of LiBeB production and the 
elemental and isotopic ratios obtained actually depend mostly on the 
ratio of the number of C and O nuclei to that of H and He.  As a 
consequence, we can conveniently specify the composition of the EPs by 
using one single parameter, $\zeta$, which we call the \emph{reduced 
metallicity} and define as:
\begin{equation}
	\zeta = \frac{^{12}\mathrm{C} + \mathrm{^{16}O}}{^{1}\mathrm{H} 
	+ \mathrm{^{4}He}},
	\label{zetaDef}
\end{equation}
where $^{12}$C, e.g., is the number abundance of the $^{12}$C nuclei 
amongst the EPs.  By way of comparison, the solar reduced metallicity 
is (with the abundances given by Anders \& Grevesse, 1989):
\begin{equation}
	\zeta_{\odot} = 1.1\,10^{-3}.
	\label{zetaSol}
\end{equation}

As can be seen from Fig.~\ref{zeta}, $\zeta_{EP}$ is about 30 to 100 
times lower than the reduced metallicity of the pure ejecta, which 
results from the dilution of the EPs with the very metal-poor ISM gas 
($Z = 10^{-4} Z_{\odot}$ for models U, and $Z = 10^{-2} Z_{\odot}$ for 
models T).  Moreover, the reduced metallicity of the EPs (at 
injection) decreases as $t^{-6/35}$, because the total SB mass 
increases as $t^{41/35}$ (see Eq.~(\ref{MSB})), while the number of 
SNe and hence the mass of the ejecta is merely proportional to time 
(constant SN power).  On the other hand, a higher ambient density and 
a higher explosion rate (SB mechanical power) both imply a smaller 
dilution, i.e.  EPs richer in C and O and thus higher spallation 
yields.

%%%%%%%%%%%%%%
\begin{figure}
\resizebox{\hsize}{!}{\includegraphics{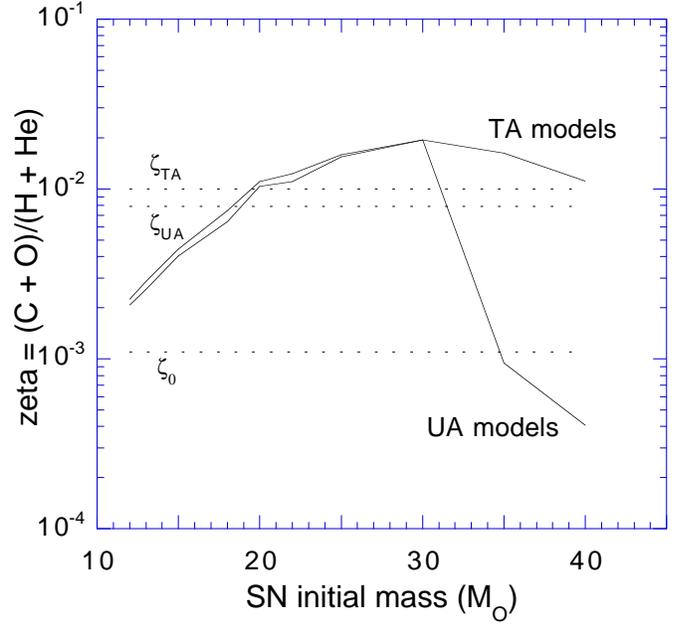}}
\caption{Reduced metallicity, $\zeta$, for different SN explosion 
models (from WW95), as a function of the initial mass of the 
progenitor.  Also shown are the solar reduced metallicity 
($\zeta_{0}$), and the mean reduced metallicity for models UA and TA.}
\label{zeta(M)}
\end{figure}
%%%%%%%%%%%%%%

Of course, the above assumption that the SN explosion rate is constant 
is certainly wrong, but as we discuss below (Sect.~\ref{Conclusion}), 
this affects the injection rate of metals and the mechanical power in 
the same way.  Reformulating the problem in terms of an `effective 
time' so that the corresponding effective SN rate is constant would 
then lead to very similar results.  However, a more subtle effect may 
lead to departures from our function $zeta(t)$ (shown in 
Fig.~\ref{zeta}): it is the fact that very massive stars eject more 
material and have a shorter life than less massive SN progenitors.  As 
a consequence, at early times of SB evolution, the mass of metals 
ejected per unit power released should be higher than at the later 
times.  Clearly, if the first star to explode has a $36\,\Msun$ 
progenitor, the induced reduced metallicity shall be two times larger 
than the value calculated above with our average model, which assumes 
that all the SNe have the same mass (i.e.  the average SN mass over 
the IMF, namely $18\,\Msun$) and the same ejecta (the mean ejecta of 
all SN progenitors).  This increase in the early value of 
$\zeta_{\mathrm{EP}}$ will of course increase the LiBeB production.  
However, it shall be compensated by a corresponding decrease at later 
times, when lower mass SNe explode, with smaller mass of metals 
ejected per unit of power released.  On the average, we don't expect 
that the total integrated yields of LiBeB should be very different 
from those calculated below, with the `average SN' model (although the 
non stationarity of the model prevents a perfect cancellation of the 
effects).

An other effect which could slightly change the results of our average 
model is the fact that the reduced metallicity of the ejecta actually 
depends on the mass of the progenitor.  This is shown in 
Fig.~\ref{zeta(M)}, for both UA and TA models.  As can be seen, 
$\zeta_{\mathrm{ej}}$ peaks for progenitors of $\sim 30\,\Msun$, where 
it is about two times higher than the average value.  Therefore, as 
above, if the first SN to explode is a $30\,\Msun$ star, we should 
expect values of $\zeta_{\mathrm{EP}}$ about two times higher than our 
average value.  Again, however, this should not affect the total LiBeB 
yields too much, as the effect will be compensated by lower 
$\zeta_{\mathrm{EP}}$ at later times.  However, it is worth noting 
that the history of the LiBeB production can be different for 
different explosion histories in the OB association (high masses first 
or low masses first), and contrary to LiBeB nucleosynthesis, this 
would have a very significant influence on the gamma-ray line emission 
rates associated with the EP interactions (to be addressed in future 
works).

\subsection{Comparison with the observations}

To compare the results of our SB model with the observations, we need 
to calculate the LiBeB as well as the Fe and O yields of the SB at the 
end of its life.  The former are the output of our time-dependent 
calculations, while the latter are taken from the SN explosion models 
of WW95, averaged over the IMF, and multiplied by the number of SNe in 
the OB association.  A successful model will then be a model which 
reproduces all the observed chemical abundance ratios.  The easiest 
way to check this is to compute the isotopic and elemental ratios of 
the light elements on the one hand, and the Be/O and Be/Fe ratios on 
the other hand.

The question of the LiBeB abundance ratios has been studied in detail 
in previous works (e.g.  Ramaty et al.  1997, Vangioni-Flam et al.  
1998, Vangioni-Flam \& Cass\'e 1999), so we focus here on the Be/O 
and Be/Fe ratios, which are the most problematic.  We only recall that 
the yields of Li, Be and B must satisfy 
$\mathcal{N}_{\mathrm{Li}}/\mathcal{N}_{\mathrm{Be}} < 100$, not to 
overproduce Li and `break the Spite plateau', and $10 \la 
\mathcal{N}_{\mathrm{B}}/\mathcal{N}_{\mathrm{Be}} \la 30$.  Both of 
these constraints are satisfied by our SB model, although the values 
which we obtain for the B/Be abundance ratio lay at the lower end of 
the above interval, i.e.  B/Be $\ga 10$.  However, it must be kept in 
mind that part of the Boron is believed to be produced by 
neutrino-spallation, during the SN explosions themselves.  This is 
indeed the only known way to get a $^{11}\mathrm{B}/^{10}\mathrm{B}$ 
isotopic ratio close to the meteoritic value of $\sim 4$, unless 
invoking a component of cosmic-rays with a very low-energy cut-off, 
which raises strong problems relating to the energetics (Ramaty et al.  
1997).  This means that the genuine B/Be production ratio in the SB, 
\emph{including the Boron produced by neutrino-spallation}, is always 
larger than the B/Be ratio derived from nuclear spallation only, which 
we calculate here.  The SB value of, say, $\mathrm{B}/\mathrm{Be} = 
11$, corresponding to a nuclear spallation isotopic ratio 
$^{11}\mathrm{B}/^{10}\mathrm{B} \sim 2.2$, implies a value of 
$\mathrm{B}/\mathrm{Be} \ga 17$ once scaled to 
$^{11}\mathrm{B}/^{10}\mathrm{B} \ga 4$, which is thus in very good 
agreement with the observations.  However, this agreement is not 
specific to the SB model, as most of the spallation models, whatever 
the EP composition and energy spectrum, would lead to essentially 
identical elemental and isotopic ratios.  As a consequence, we shall 
pay more attention to the Be production as compared to O and Fe in the 
following, because this represents the most conclusive argument in 
favour of the SB model.

%%%%%%%%%%%%%%
\begin{figure*}
\centerline{\psfig{file=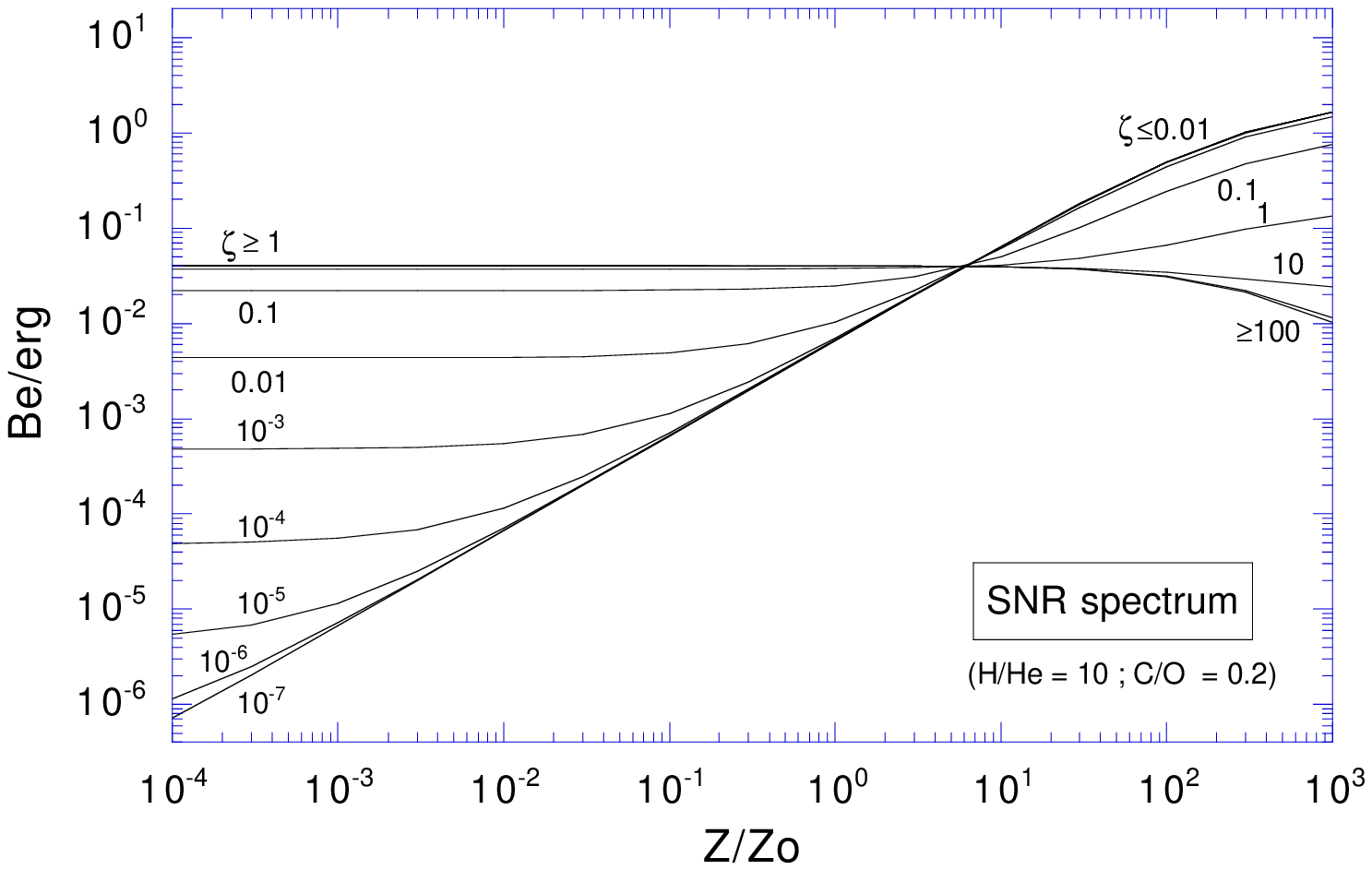, width=8.5cm}
            \hfill
            \psfig{file=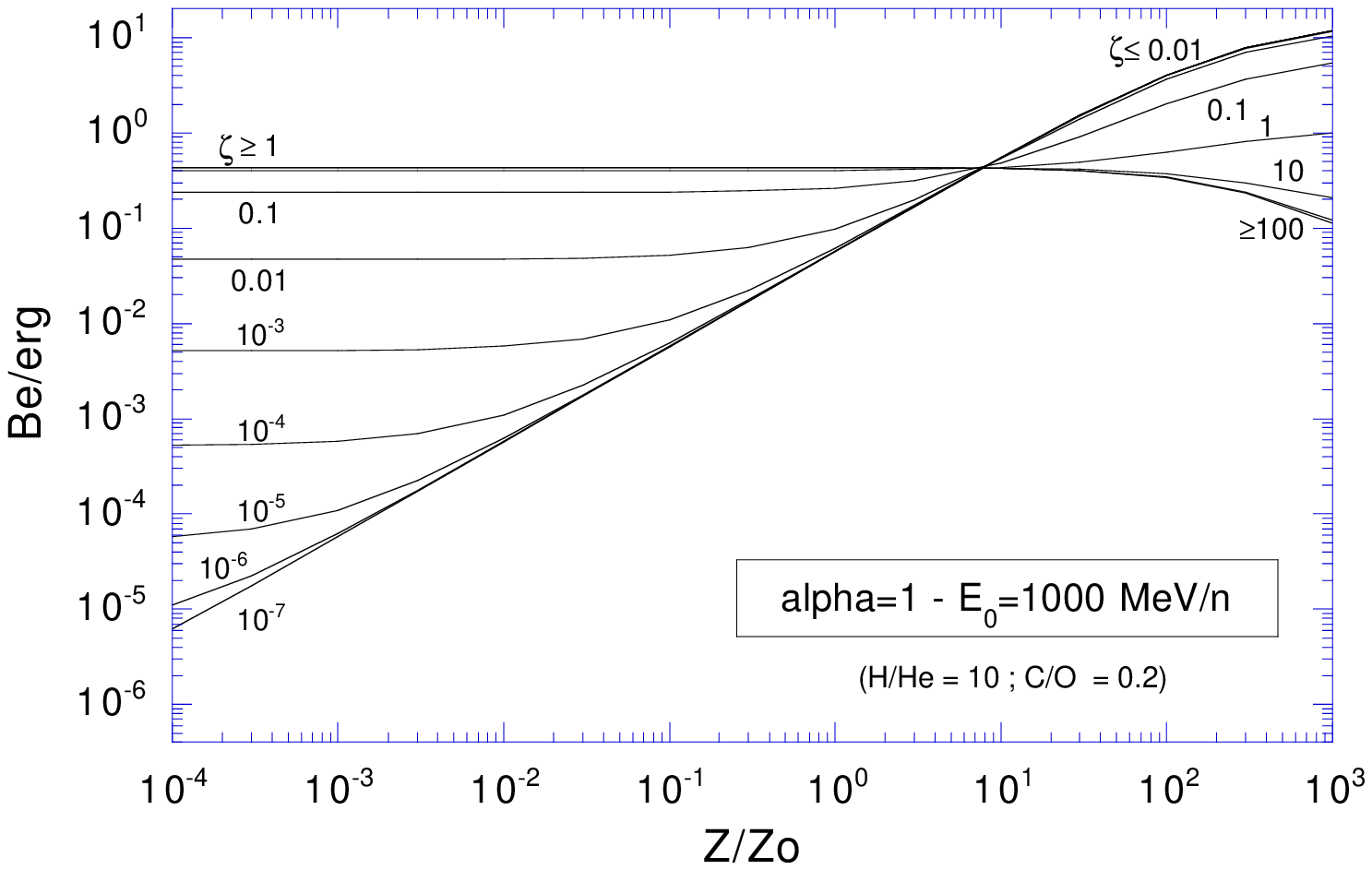, width=8.5cm}}
\centerline{\psfig{file=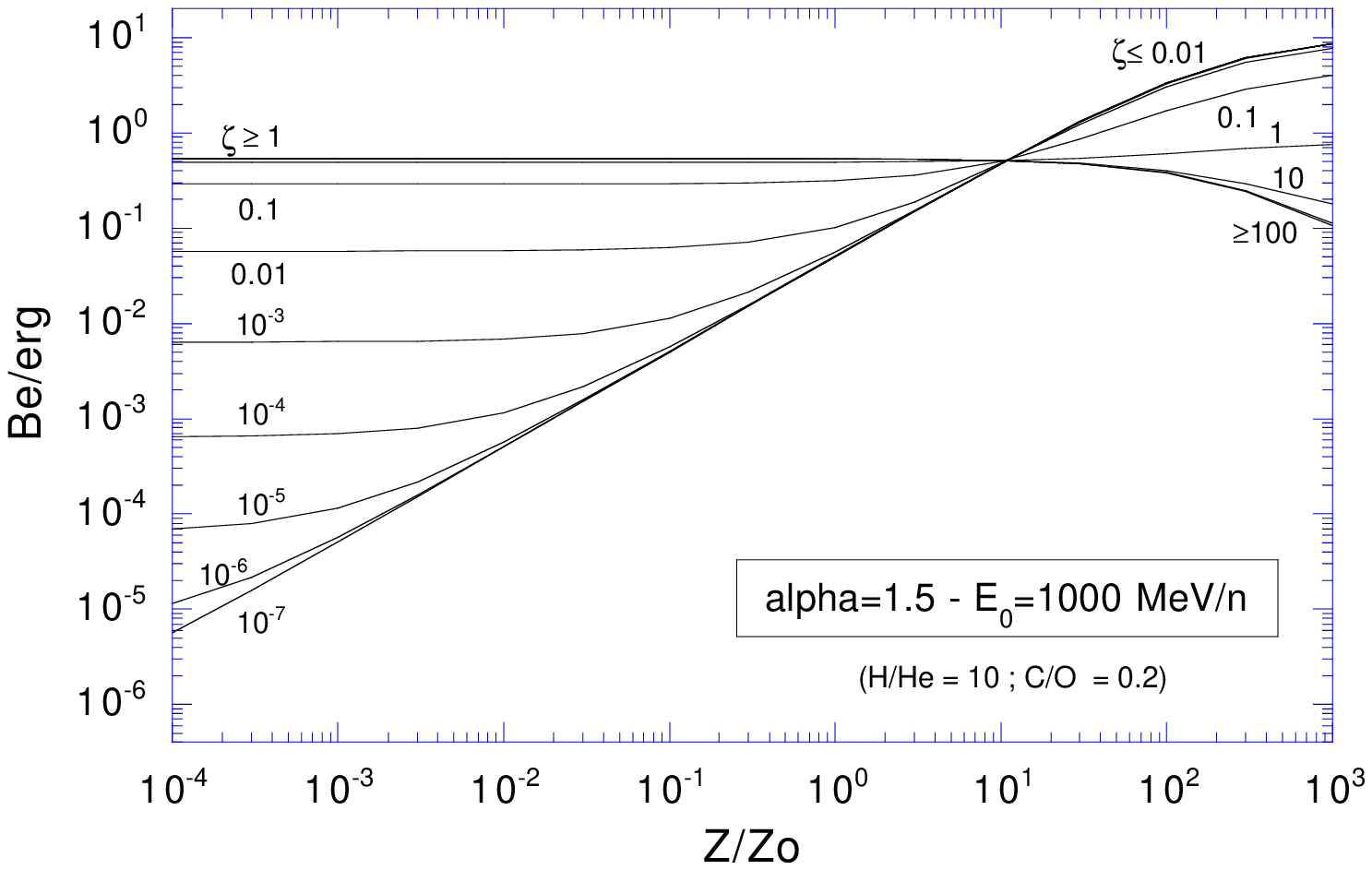, width=8.5cm}
            \hfill
            \psfig{file=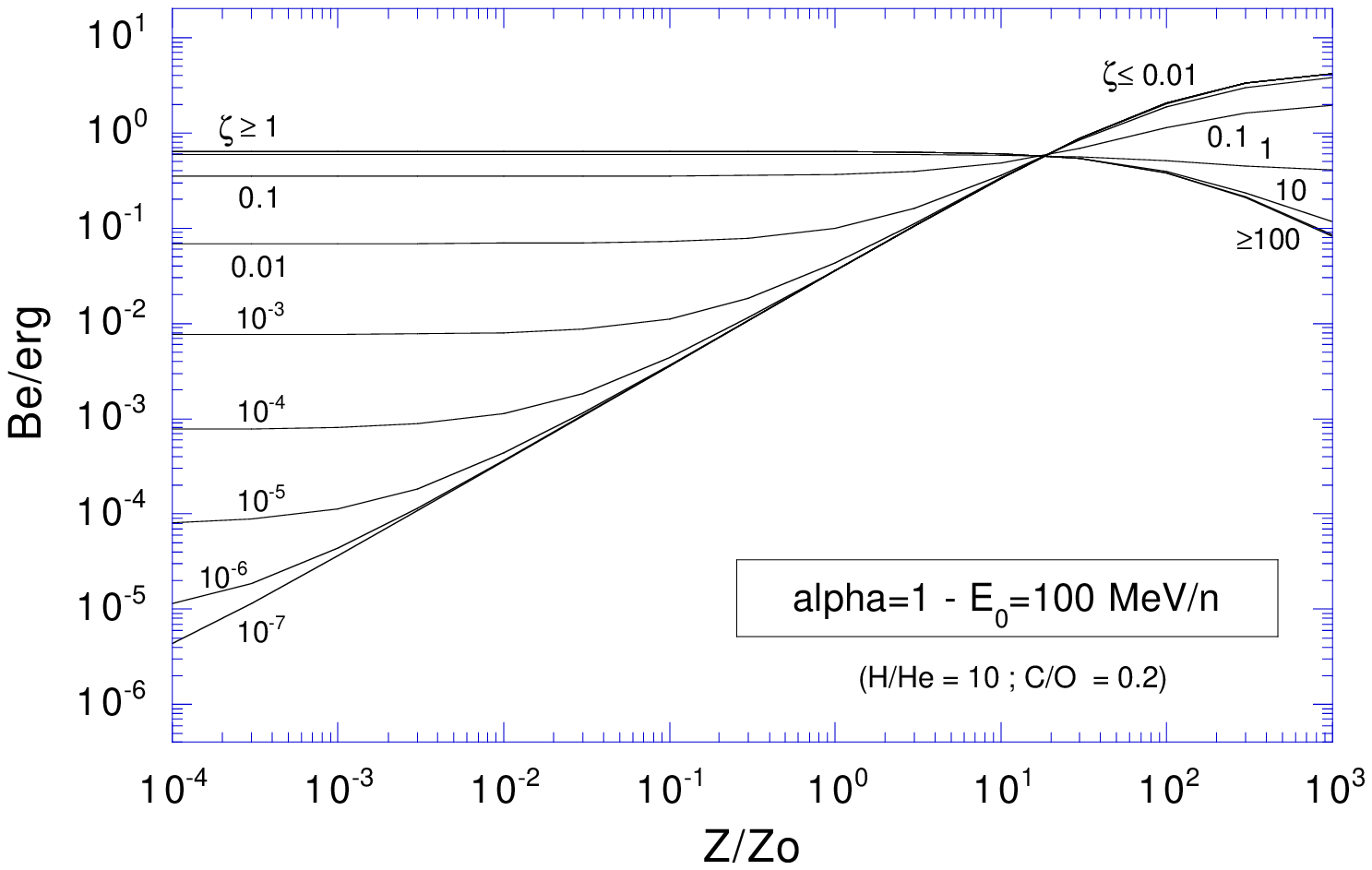, width=8.5cm}}
\caption{Be production efficiency by spallation, in numbers of Be 
nuclei produced per erg of EPs injected in the ISM, for different 
energy spectra and EP compositions (as indicated on the figures), as a 
function of the ISM (target) metallicity, expressed in units of the 
solar metallicity.  The labels refer to the reduced metallicity of the 
EPs, $\zeta$, defined in the text.}
\label{Be/erg}
\end{figure*}
%%%%%%%%%%%%%%

%%%%%%%%%%%%%%
\begin{figure*}
\centerline{\psfig{file=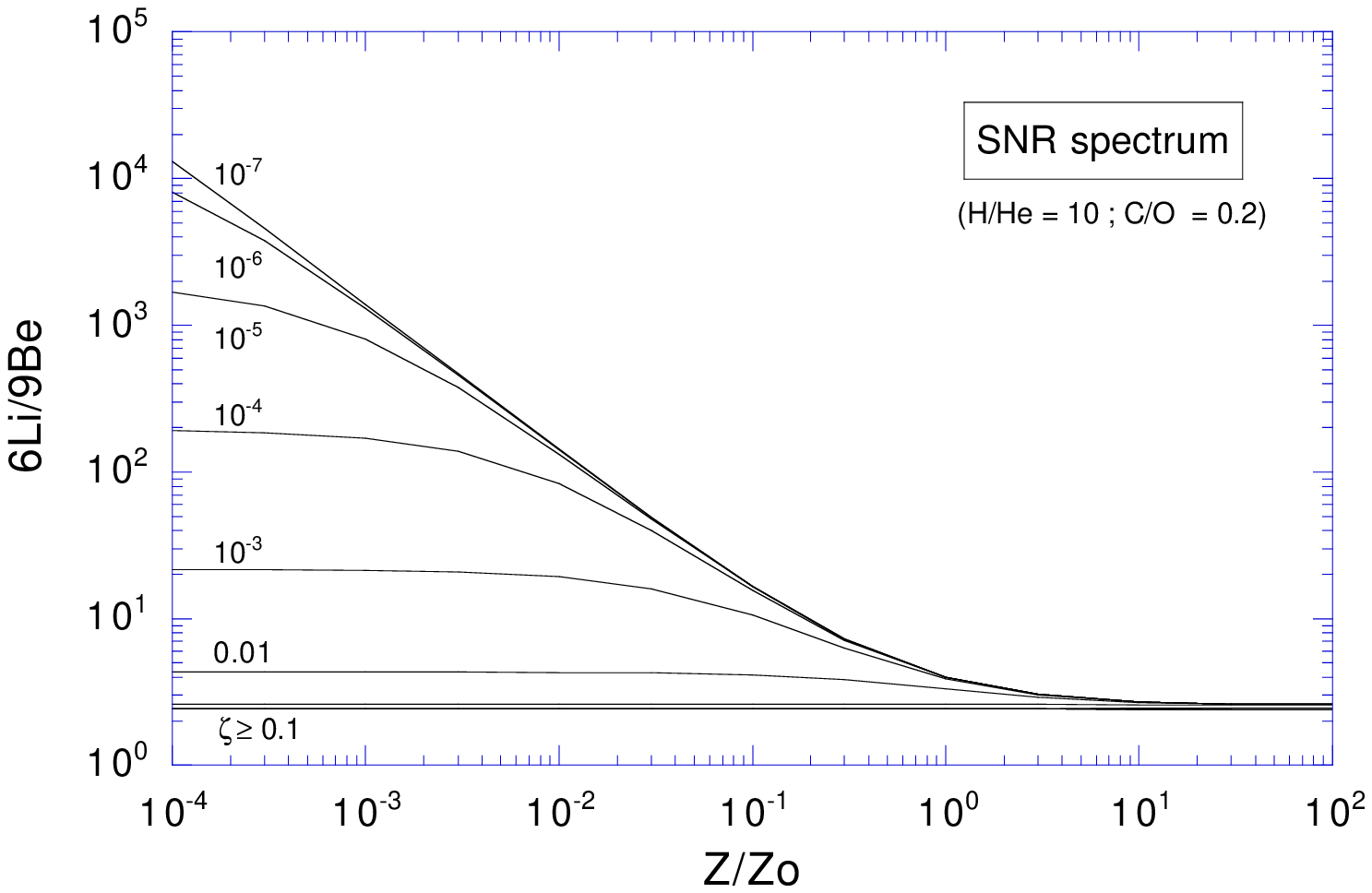, width=8.5cm}
            \hfill
            \psfig{file=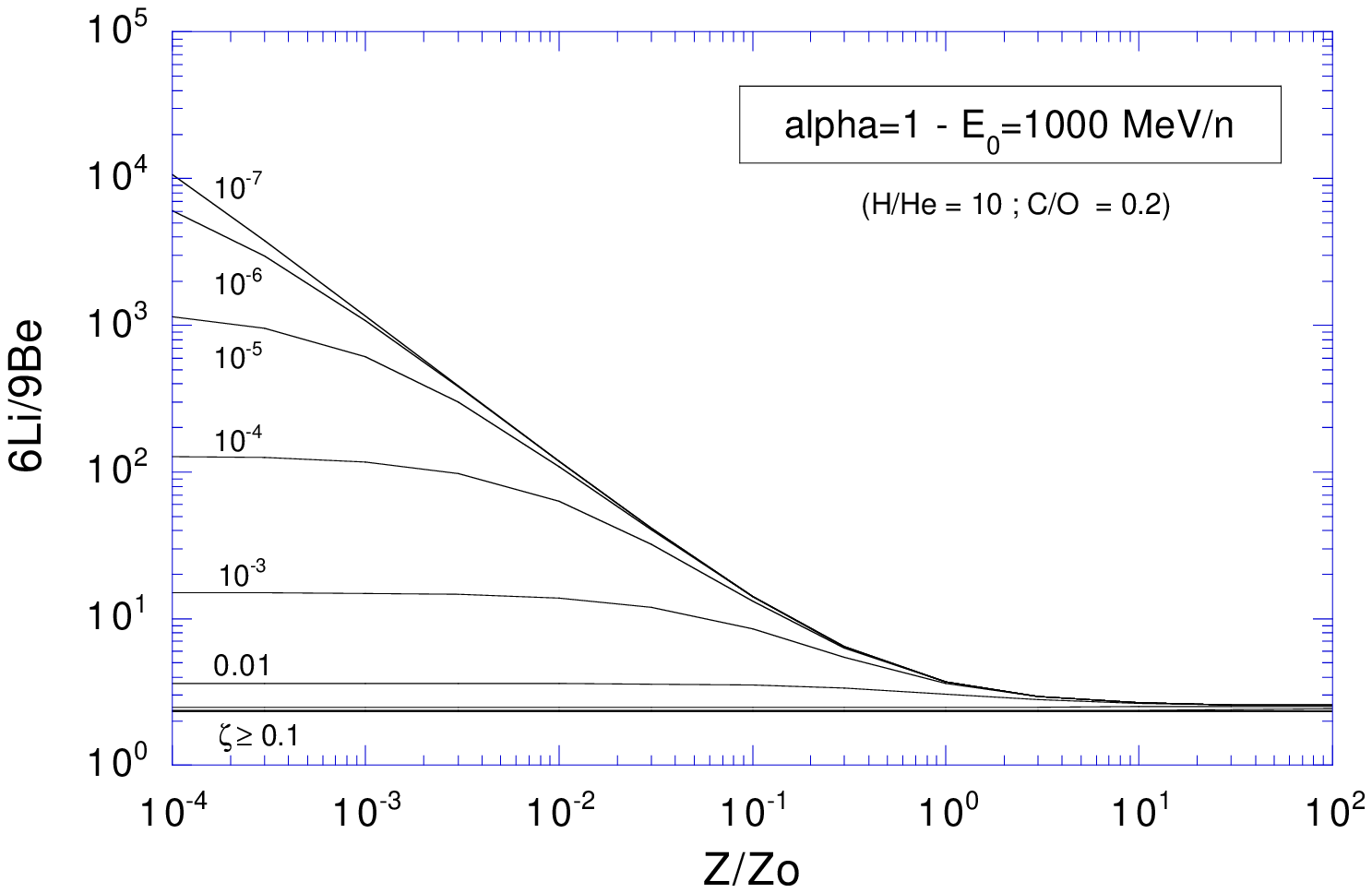, width=8.5cm}}
\centerline{\psfig{file=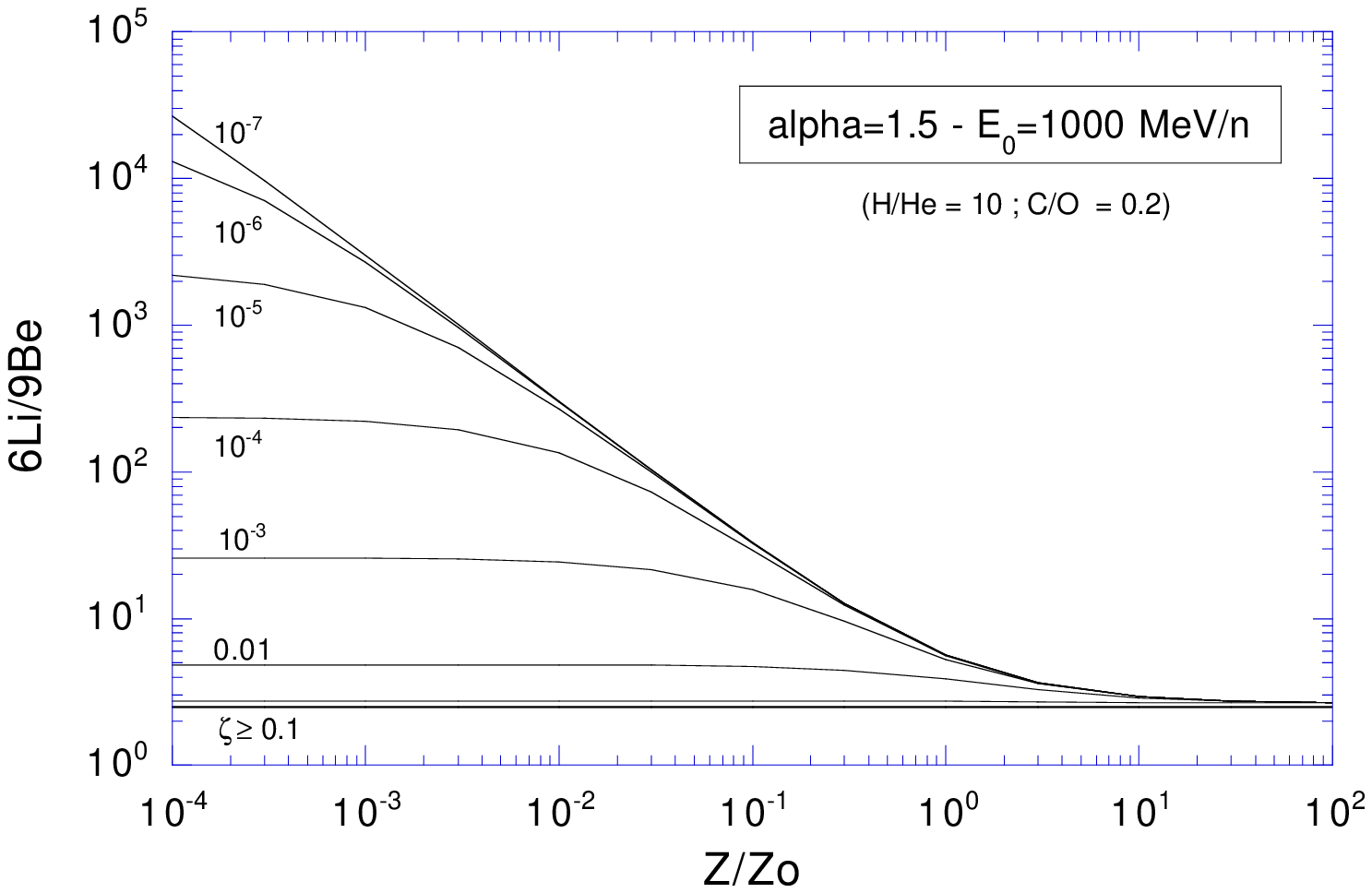, width=8.5cm}
            \hfill
            \psfig{file=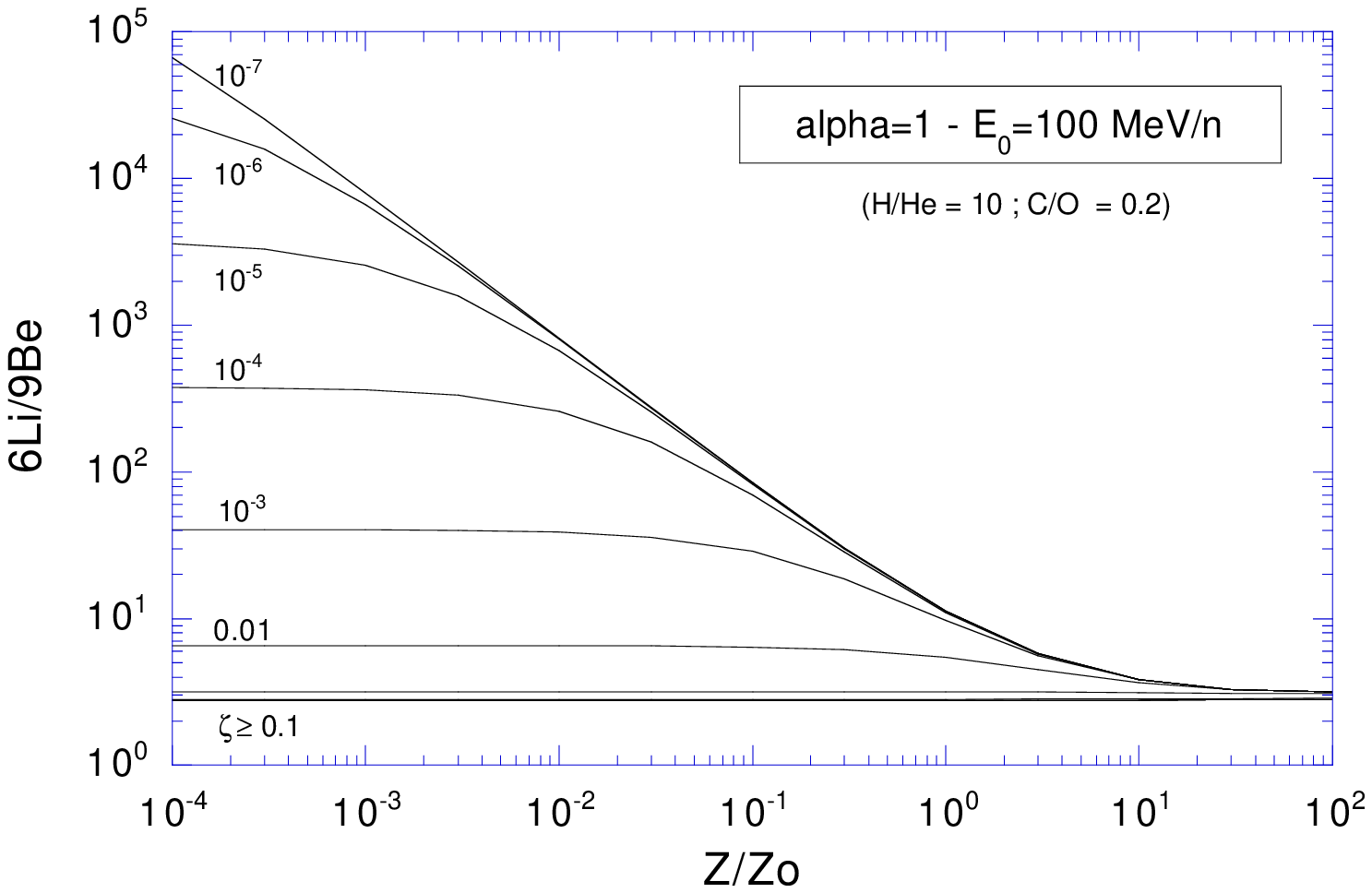, width=8.5cm}}
\caption{Production ratio of $^{6}$Li and $^{9}$Be by spallation, for 
different EP spectra and compositions, and different target 
compositions, as in Fig.~\ref{Be/erg}.}
\label{6/9}
\end{figure*}
%%%%%%%%%%%%%%

To have done with the LiBeB abundance ratios, we note that recent 
observational work also provided the first measurements of the 
$^{6}$Li isotope in two halo stars of metallicity $Z \simeq 
10^{-2.3}Z_{\odot}$ (Hobbs \& Thorburn, 1994, 1997; Smith et al., 
1998; Cayrel et al., 1999).  The deduced $^{6}$Li/$^{9}$Be ratio in 
these stars is found to lay between 20 and 80 (although Be has not 
been measured directly in these particular stars), in contrast with 
the solar value of $\sim 6$.

Concerning the Be/Fe ratio, the observations show that it is 
approximately constant for stars of metallicity up to $10^{-1}$ solar, 
and equal to $\sim 1.6\,10^{-6}$ (e.g.  Ramaty et al.  1997).  This is 
the value which the SNR model failed to obtain (Parizot \& Drury, 
1999a,b), and the most constraining of all the observational 
requirements, in the present state of knowledge.  Assuming an average 
Fe yield of $0.11~\Msun$ per SN, this requires a production of $\sim 
3.8\,10^{48}$ Be nuclei per SN. As for the Be/O ratio, the situation 
is less clear from the observational point of view, because there are 
no data at metallicities below $10^{-2}Z_{\odot}$.  However, since the 
SB model predicts a constant Be/O ratio at low metallicity, we assume 
that this is actually the case, and derive a value of $\sim 
7.5\,10^{-5}$ from the available observational data (e.g.  Fields \& 
Olive, 1999).

Interestingly enough, the above Be/Fe and Be/O ratios are not mutually 
compatible, considering the Fe and O yields given by the SN models of 
WW95.  This is due to the failure of these models to reproduce 
consistently the O and Fe evolution in halo stars, as for instance the 
observed increase of the O/Fe ratio towards the lowest metallicities 
(Israelian et al.  1998, Boesgaard et al.  1998), although a different 
mixing time of freshly synthesized Fe and O nuclei with the ISM gas 
may solve this problem (Ramaty \& Lingenfelter, 1999).  As far as the 
LiBeB production is concerned, however, the most relevant ratio to be 
compared with the observations is Be/O, as Oxygen is the main 
progenitor of Be, while the genuine Fe production in SNe has no 
influence at all on the Be yield of the SB. A simple solution to the 
problem might thus simply be that the SN models produce too much Fe at 
low metallicity, due possibly to a wrong determination of the mass 
cut-off (see Parizot \& Drury, 1999c, for a discussion of this 
issue).

\section{Results, analysis and discussion}

\subsection{General results from steady-state calculations}
\label{Steady}

Before we present the results of the SB model, it is worthwhile 
showing some general results obtained with steady-state calculations, 
for different EP spectra and compositions, as a function of the target 
metallicity.  As already mentioned, H, He, C and O are the only 
relevant elements, so we use the reduced metallicity $\zeta$ defined 
above instead of the usual $Z$.

In Fig.~\ref{Be/erg}, we plot the Be production efficiency, defined as 
the number of Be nuclei synthesized per erg injected in the form of 
EPs, for different spectra, namely the cosmic-ray source spectrum and 
the `SB spectrum' defined above, with spectral index $\alpha = 1.5$ or 
$\alpha = 1$.  In the latter case, two different cut-off energies have 
been used, as indicated on the plots.  Finally, for each spectrum, we 
calculated the Be production efficiency for eleven values of the EP 
reduced metallicity, ranging from $10^{-7}$ to $10^{3}$ (recalling 
that $\zeta_{\odot} = 1.1\,10^{-3}$).  These curves enable one to 
quickly evaluate the Be production for any source and target 
composition, as well as for the most commonly used EP spectra.  Their 
qualitative analysis is straightforward, and independent of the EP 
spectrum.

In the upper-left part of the figures, the Be production efficiency is 
an horizontal line for any given $\zeta_{\mathrm{EP}}$, which 
indicates that the inverse spallation reactions dominate.  As expected 
in this case, the production efficiency is independent of the target 
metallicity, whence the horizontal lines.  Of course, this regime 
dominated by inverse spallations ends up at lower target metallicities 
for lower $\zeta_{\mathrm{EP}}$, as can be seen on Fig.~\ref{Be/erg}.  
Clearly visible on the figures is also the straight line with slope~1 
where all the curves converge.  This line corresponds to a Be 
production dominated by direct spallation reactions.  This roughly 
amounts to say that the target metallicity is higher than the EP 
metallicity, so that the production efficiency is virtually 
independent of the EP composition.  The reactions producing Be are of 
the type $\mathrm{p,}\alpha + \mathrm{C,O}$, and the production 
efficiency is just proportional to the target metallicity.  In the 
upper-left part of the figures, where inverse spallation dominates, 
the opposite holds: the production efficiency is proportional to the 
EP (reduced) metallicity, except for very high $\zeta$, namely 
$\zeta\ga 0.1 \simeq 100\, \zeta_{\odot}$, where the process 
saturates.  The corresponding value is thus the highest production 
efficiency which one can get with the particular spectrum considered.  
Increasing the target metallicity then brings all the curves together 
towards one particular point where the production efficiency is the 
same, whatever the EP composition.  This `meeting point' is seen to 
correspond to a target metallicity between 60 and 200 times solar.  
Above this point, the Be production efficiency is actually higher for 
lower EP metallicities.  This results from the fact that in a very 
rich target, energetic H and He nuclei are more efficient than C and 
O, because at a given energy per nucleon, they carry much less energy.

This behavior of the production efficiency is exactly the same for B, 
although with values about 11 times higher (i.e.  B/Be $\sim 11$).  We 
therefore do not show these curves here.

As for Li, we show on Fig.~\ref{6/9} the evolution of the 
$^{6}$Li/$^{9}$Be ratio under the same conditions as above.  To 
understand the curves, one only needs to know that $^{6}$Li can be 
produced either by fusion reactions ($\alpha + \alpha$) or by 
spallation reaction implying C, N or O nuclei, while $^{9}$Be can only 
be obtained by the latter processes.  As a consequence, the 
$^{6}$Li/$^{9}$Be production ratio reflects the 
$\mathrm{He}/(\mathrm{C} + \mathrm{O})$ ratio either in the EPs or in 
the target.  Two main regimes can be observed on Fig.~\ref{6/9}, 
depending on whether fusion or spallation reactions dominate the 
$^{6}$Li production.  For high enough EP metallicities 
($\zeta_{\mathrm{EP}}\ga 0.1$), spallation dominates, so that both 
$^{6}$Li and $^{9}$Be are produced in the same way.  Their production 
ratio is thus essentially constant, determined by the ratio of the 
cross-sections (averaged over the spectrum).  For lower EP 
metallicities, however, the $^{6}$Li/$^{9}$Be production ratio is 
proportional to the fusion/spallation production ratio of Li, or to 
the $\mathrm{He}/(\mathrm{C} + \mathrm{O})$ ratio, that is practically 
to $\zeta_{\mathrm{EP}}^{-1}$, as seen on Fig.~\ref{6/9}.  Finally, 
the same behavior as for Be relative to the target metallicity can 
still be observed: at low $Z$, the $^{6}$Li/$^{9}$Be production ratio 
is independent of $Z$, because the inverse spallation reactions 
dominate, while when $\zeta_{\mathrm{target}} \ga 
\zeta_{\mathrm{EP}}$, $^{6}$Li/$^{9}$Be is inversely proportional to 
$\zeta_{\mathrm{target}}$, because the direct spallation dominates 
(i.e.  the $^{6}$Li production efficiency is constant while $^{9}$Be's 
is proportional to $\zeta_{\mathrm{target}}$).

Coming now to the quantitative point of view, it turns out that 
Figs.~\ref{Be/erg} and~\ref{6/9} jointly provide important clues 
towards a solution of the \libeb~evolution problem in the Galaxy.  As 
mentioned earlier, the observationally required Be production per SN 
is of order $4\,10^{48}$, if comparison is made with Fe, or $\sim 
6\,10^{47}$, if comparison is made with O (assuming the averaged 
Oxygen yield of 1.1~\Msun per SN, derived from the models of WW95).  
If the mean SN explosion energy of $10^{51}$~erg and the acceleration 
efficiency is 10~\%, this implies a Be production efficiency of $\sim
4\,10^{-2}$ Be/erg (for comparison with Fe), or $\sim 6\,10^{-3}$ 
Be/erg (for comparison with O), the latter being more reliable, as 
discussed above.  Reporting to Fig.~\ref{Be/erg}, it is easy to 
determine the EP reduced metallicity which is needed to get such an 
efficiency.  Since we are interested in low-metallicity targets (early 
Galaxy), the correspondence between the Be production efficiency and 
$\zeta_{\mathrm{EP}}$ is one to one, once a spectrum is given.  For a 
cosmic-ray source spectrum (Fig.~\ref{Be/erg}a), an efficiency of 
$6\,10^{-3}$ Be/erg requires an EP reduced metallicity about 20 times 
the solar value.  This seems quite hard to achieve, even in the SB 
model, since Fig.~\ref{zeta} shows values between 50 and 200 times 
lower.  On the other hand, a SB spectrum with a relatively low-energy 
cut-off can achieve the required Be production efficiency with values 
of $\zeta_{\mathrm{EP}}$ as low as a few $10^{-4}$, i.e.  close to the 
values obtained naturally within the framework of our SB model.

Further referring to Fig.~\ref{6/9}, it is quite interesting to see 
that our SB values for $\zeta_{\mathrm{EP}}$ of a few $10^{-4}$ also 
predict a $^{6}$Li/$^{9}$Be production ratio decreasing from a few 
tens in low-metallicity targets to about 6 around solar metallicity.  
This is exactly the observed behavior, although more data on the 
$^{6}$Li abundance in low-metallicity stars are still needed to draw a 
definite conclusion.  We should also mention that Vangioni-Flam et al.  
(1999) and Fields \& Olive (1999) proposed different models to 
account for the decrease of $^{6}$Li/$^{9}$Be with metallicity.  In 
any case, however, the above qualitative and quantitative analysis 
based on steady-state calculations show that the SB model is a very 
good candidate to solve the LiBeB evolution problem in the early 
Galaxy, by naturally predicting an EP composition having the 
appropriate reduced metallicity to provide both the correct Be 
production efficiency and the correct $^{6}$Li/$^{9}$Be ratio.  We now 
show that this model indeed works when applied more accurately using 
time-dependent calculations.

\subsection{Time-dependent calculations for the SB model}

The SB model has been described above in Sect.~\ref{Model}.  The main 
parameters are the mechanical power, $\mathcal{P}_{\mathrm{in}}$, 
provided by the OB association to the superbubble 
($\mathcal{P}_{\mathrm{in}} = 10^{38}$~erg for all the plots shown 
here), the lifetime of the SB, $\tau_{\mathrm{SB}}$, the ambient ISM 
density, $n_{0}$, the SN explosion model (UA or TA, from WW95), the EP 
spectrum (SNR or SB spectrum, with index $\alpha = 1$ or 1.5), and the 
cut-off energy, $E_{0}$, of the spectrum (except for the SNR spectrum, 
where it is assumed constant at $10^{14}$~eV/n).

%%%%%%%%%%%%%%
\begin{figure}
\resizebox{\hsize}{!}{\includegraphics{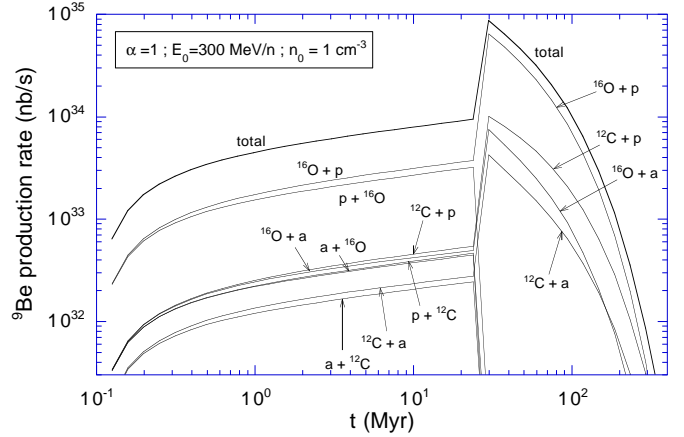}} 
\caption{Detailed Be production rate by the SB model, as a function of 
time.  The EP spectrum is a SB spectrum with spectral index $\alpha = 
1$ and a cut-off energy of $E_{0} = 300$~MeV/n, as indicated.  The 
ambient ISM density is 1 cm$^{-3}$.  The labels of the curves refer to 
the corresponding spallation reaction, where the first nucleus 
mentioned is the energetic one.}
\label{Be-detail}
\end{figure}
%%%%%%%%%%%%%%

In Fig.~\ref{Be-detail}, we show the detailed Be production rate as a 
function of time for a SB spectrum with $\alpha = 1$ and $E_{0} = 
300$~MeV/n, and an ISM density of $1\,\mathrm{cm}^{-3}$.  As expected, 
the main producing reaction is the spallation of $^{16}$O by the H 
nuclei, because O is the most abundant metal in the SN ejecta.  The 
total production rate is shown to increase during the whole lifetime 
of the SB (here $\tau_{\mathrm{SB}} = 30$~Myr), as a result of the 
following competing effects: i) the accumulation of the EPs in the SB, 
ii) the decrease of the EP and target metallicity ($\propto 
t^{-6/35}$), iii) the Coulombian energy loss of the EPs, and the 
rarefaction of the gas inside the expanding SB. Neglecting the energy 
losses (which act on a longer timescale, because of the low SB 
density), it is easy to estimate the time dependence of the Be 
production rate:

\begin{equation}
	\frac{\d\mathcal{N}_{\mathrm{Be}}}{\d t} \propto
	\mathcal{N}_{\mathrm{EP}}\alpha_{\mathrm{O}}n_{\mathrm{SB}},
	\label{BeProdRate}
\end{equation}
where $\mathcal{N}_{\mathrm{EP}}$ is the total number of EPs in the SB 
and $\alpha_{\mathrm{O}}$ is the fraction of Oxygen in the EPs 
(inverse spallation) or the SB gas (direct spallation).

Now $\mathcal{N}_{\mathrm{EP}}\propto t$, because the acceleration 
rate is constant, and $\alpha_{\mathrm{O}}\propto t^{-6/35}$, as 
already mentioned.  Considering the time dependence of 
$n_{\mathrm{SB}}$ (Eq.~\ref{nSB}), we thus obtain 
$\d\mathcal{N}_{\mathrm{Be}}/\d t\propto t^{1/5}$, in very good 
agreement with the $\d\mathcal{N}_{\mathrm{Be}}/\d t\propto t^{0.22}$ 
obtained from Fig.~\ref{BeProdRate} by a fit between 1 and 10~Myr.  
Note finally that, at a given time, the fraction of Oxygen, 
$\alpha_{\mathrm{O}}$ is actually slightly higher in the EPs than in 
the SB, because the EPs collect particles accelerated at earlier times 
when the metallicity was higher, which explains why the inverse 
spallation reaction $^{16}\mathrm{O} + \mathrm{p}$ actually dominates, 
as seen on Fig.~\ref{Be-detail}.

As expected, Fig.~\ref{Be-detail} also shows the (discontinuous) 
increase of the production rates at $\tau_{\mathrm{SB}}$, when the EPs 
leave the SB and interact with a much denser medium.  The situation 
has been idealized here, as we should not expect that the EPs leave 
the SB instantaneously at $\tau_{\mathrm{SB}}$, nor that they be 
`perfectly confined' within the SB before $\tau_{\mathrm{SB}}$.  
However, as we emphasized above, this idealized scenario does provide 
the correct \emph{total} Be yield, while giving a valuable insight on 
the respective role of direct and inverse spallation reactions.  For 
example, it is evident that only the latter are efficient in the low 
metallicity target relevant at $t > \tau_{\mathrm{SB}}$.  The nuclear 
reaction rates are also seen to decrease after $\tau_{\mathrm{SB}}$, 
due to the Coulombian energy losses in the higher density medium.  As 
mentioned at the end of Sect.~\ref{LiBeBInSBs}, we arbitrarily chose 
the target density to be 0.1~cm$^{-3}$ after $\tau_{\mathrm{SB}}$.  
This choice has no influence on the total Be yield, since once the 
acceleration has stopped, the integral of the reaction rates is 
independent of density (see Parizot \& Drury, 1999a, for more 
details).

%%%%%%%%%%%%%%
\begin{figure}
\resizebox{\hsize}{!}{\includegraphics{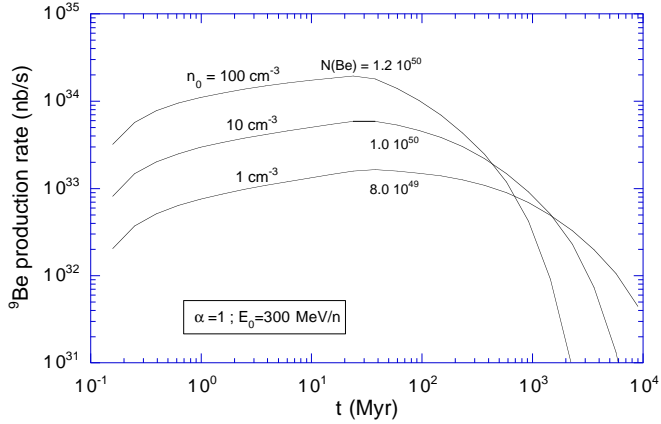}}
\caption{Be production rate (in number of nuclei per second) as a 
function of time for the same SB model as in Fig.~\ref{BeProdRate}, 
but for different ISM densities.  The total, integrated Be yield is 
indicated in each case, beside the corresponding curve.}
\label{Be(n0)}
\end{figure}
%%%%%%%%%%%%%%

This is no longer true, however, when the dynamics of the SB evolution 
is involved (Parizot 1999).  This is shown on Fig.~\ref{Be(n0)}, where 
the Be production rate is plot as a function of time, for different 
values of the ambient density.  For $t < \tau_{\mathrm{SB}}$, the 
reaction rates are proportional to the SB density, that is to 
$n_{0}^{19/35}$, while after $\tau_{\mathrm{SB}}$, the EPs lose energy 
on a time scale all the shorter that the density is high.  Concerning 
the total, integrated yields of Be, they are also indicated on 
Fig.~\ref{Be(n0)}.  They show only a weak dependence on the ambient 
density, which results from the fact that most of the Be nuclei are 
produced by EP interactions outside the SB, or in the dense shell of 
interstellar gas.  Note that we have used here a different 
prescription for the target density after $\tau_{\mathrm{SB}}$.  It 
has been choosen so as to ensure the continuity of the production 
rates.

%%%%%%%%%%%%%%
\begin{figure}
\resizebox{\hsize}{!}{\includegraphics{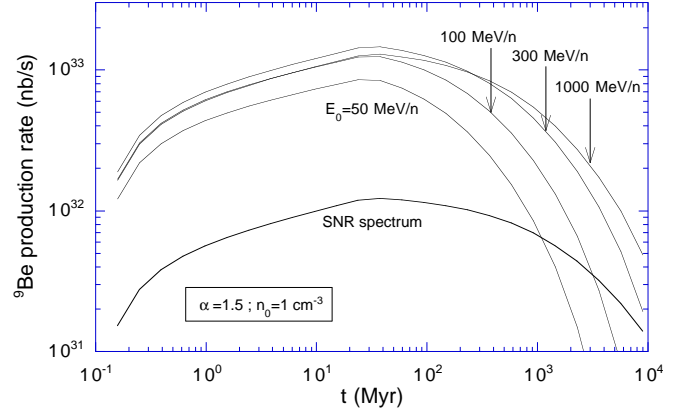}}
\caption{Be production rate as a function of time, for SB models using 
the SB spectrum with index $\alpha = 1.5$ and different cut-off 
energies. The case of a SNR source spectrum is also shown.}
\label{Be(E0)}
\end{figure}
%%%%%%%%%%%%%%

The same has been used in Fig.~\ref{Be(E0)}, where we show the Be 
production rate as a function of time for a SB spectrum with index 
$\alpha = 1.5$, for different values of the cut-off energy.  We also 
show, for the sake of comparison, the Be production rate obtained with 
the SNR spectrum.  It is clearly less efficient, as expected from the 
steady-state analysis above, although the production rates decrease 
less quickly after $\tau_{\mathrm{SB}}$, because of the high energy 
particles, which keep supernuclear energies (i.e.  above nuclear 
reaction thresholds) for a longer time.

\begin{figure}
\resizebox{\hsize}{!}{\includegraphics{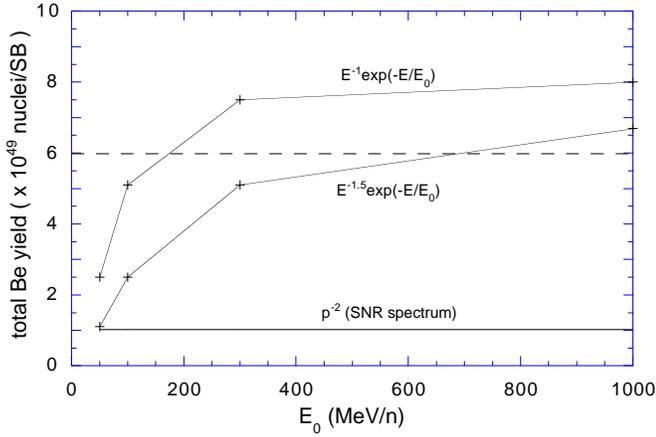}}
\caption{Total Be yields of the SB for different EP spectra, as a 
function of the cut-off energy, $E_{0}$. For the SNR spectrum, the 
cut-off energy is irrelevant, and we show a constant yield. The dashed 
line corresponds to the yield required to explain the observed Be/O 
ratio (see text). In this SB model, a total of 100 SNe have exploded during 
the SB lifetime of 30 Myr.}
\label{BeYields}
\end{figure}

The most important results relate to the total Be yield, 
$\mathcal{N}_{\mathrm{Be}}$, obtained with the SB model.  This is 
obtained by merely integrating the production rates over time.  The 
results are shown on Fig.~\ref{BeYields}, for different spectra, as a 
function of the cut-off energy, $E_{0}$.  The horizontal dashed line 
corresponds to the observed value of the Be/O ratio, derived as 
explained above (in this model, the SB is blown by 100 SNe, which must 
be accompanied by the production of $\sim 6\,10^{49}$ of Be).  As can 
be seen, the results of our SB model are in very good agreement with 
the observations, if the EP spectrum is of the SB type, with a cut-off 
energy of a few hundreds of MeV/n.  Interestingly enough, this is just 
the energy range predicted by the SB acceleration model (Bykov \& 
Fleishman, 1992; Bykov, 1995,1999).  On the other hand, the standard 
SNR spectrum leads to a Be yield a factor of 6 too low, which 
rules out this spectrum unless strongly selective acceleration is 
invoked.

An enhancement of the C and O abundance relative to H and He by a 
factor of 6 does not seem unrealistic at first glance.  Indeed, the 
study of the CR composition indicates that a similar enhancement by a 
factor of $\sim 8$ must have occurred if the cosmic rays are 
accelerated out of the ISM. Moreover, such an enhancement is 
satisfactorily accounted for by the theoretical CR acceleration model 
of Ellison et al.  (1997), where refractory elements are locked in 
grains and preferentially accelerated, while the acceleration of 
volatile elements (like H and He) is less efficient.  However, the 
selectivity of this acceleration mechanism depends to a large extent 
on the physical conditions in the ambient medium, notably the 
temperature, which determines the degree of ionization of the 
different nuclei, and the fraction of the elements which are locked 
into grains.  Therefore, it is not clear whether the specific 
conditions prevailing in a SB give rise to the same enhancement of C 
and O as in the usual ISM. Additional work would thus be needed to 
determine whether a selective acceleration mechanism increasing 
$\zeta_{EP}$ by a factor of~6 in the SB is reasonable or not.

%%%%%%%%%%%%%%
%%%%%%%%%%%%%%

As it stands, two opposite positions can be adopted within the 
framework of the SB model: i) holding on to the SB acceleration 
mechanism of Bykov et al.  and thus to the SB spectrum, because it 
does not require any selective acceleration, or ii) preferring to 
invoke selective acceleration and keep the usual SNR (cosmic ray 
source) spectrum, because it does not require a different acceleration 
mechanism.  Both points of view seem to be acceptable in the current 
state of knowledge, and provide genuine solutions to the LiBeB 
evolution problem in the early Galaxy, as we have shown.  Additional 
theoretical work, relating to the SB acceleration mechanism on the one 
hand, and to the selectivity of the CR acceleration mechanism under 
the physical conditions prevailing in a SB, on the other hand, is now 
needed to decide between these two stances.

Interestingly enough, a full solution of the \libeb~problem therefore 
appears to be able to provide important information about related 
fields in high energy astrophysics.  For instance, one important 
remaining questioning about energetic particles in the Galaxy concerns 
the part of the spectrum below 1~GeV/n.  Since these so-called low 
energy cosmic rays are prevented from reaching the inner solar system 
by the magnetic fields associated with the solar wind, we cannot get 
any direct information about their spectrum and composition.  While it 
may seem reasonable to assume that these are similar to the ordinary 
cosmic rays (CRs), it remains perfectly possible that a second 
component of EPs, with different origin, spectrum and composition, 
also exists at low energy ($E\la 1$~GeV/n) and is superimposed to the 
ordinary CRs.  Now since virtually all the spallative LiBeB nuclei are 
expected to be produced by these low energy cosmic rays, whatever 
their spectrum and composition, it is clear that strong constraints on 
their characteristics should come out of a detailed study of the light 
element evolution in the Galaxy.  The results presented here seem to 
indicate that the SBs could well be the source of an important 
component of EPs, dominating the usual CRs at low energy, and 
therefore being responsible for most of the LiBeB production in the 
Galaxy.

\section{Conclusion}
\label{Conclusion}

In this paper we have shown that, even in the very early Galaxy, 
superbubbles are natural sources of C and O-rich energetic particles 
with a metallicity of about $10^{-1}Z_{\odot}$, even within the most 
conservative assumption concerning the mixing and dilution of the SN 
material ejected inside the SB with the ISM gas evaporated of the 
expanding SB shell and/or embedded clouds.  We further showed that 
these enriched EPs have enough energy to produce as much Li, B and B 
as observed in the metal-poor halo stars.  This SB model thus provides 
a solution of the long-standing problem of LiBeB evolution in the 
early Galaxy.  In addition, it offers a straightforward way to account 
for the most recent $^{6}$Li data at low metallicity, as discussed in 
Sect.~\ref{Steady}.

From our point of view, the most tantalizing solution seems to be the 
SB model using the so-called SB spectrum, as derived from the SB 
acceleration model of Bykov et al., since it does not require any ad 
hoc assumption concerning the mixing of the SN ejecta and the chemical 
selectivity of the acceleration mechanism.  However, from another 
point of view, it could be argued that invoking a selective 
acceleration and/or a very imperfect mixing of the gas inside the SB 
is preferable, because it avoids the recourse to an energy spectrum 
different from the cosmic rays.

One of the main original features of our model is that it implies a 
decoupling between the metallicity of the stars and their age, or in 
other words, a non monotonic Galactic enrichment.  In particular, 
stars formed at the same time can have metallicities spreading over 
two orders of magnitude or more.  Indeed, as discussed in 
Sect.~\ref{Overview}, a SB developing in the early Galaxy within an 
medium of metallicity, say, $10^{-4}Z_{\odot}$, provides $\la 
10^{5}\,\Msun$ of gas enriched to $\sim 10^{-1}Z_{\odot}$, with the 
observed Be/O and LiBeB abundance ratios.  This gas then mixes with 
the ambient low-metallicity gas and collapses to form stars of various 
metallicities, ranging from, say, $10^{-4}$ to $10^{-1}Z_{\odot}$.  
Evidently, all these stars show the same abundance ratios, and the 
linear increase of the Be and B abundances with the stellar 
metallicity is, in this model, the manifestation of a \emph{dilution 
line}.  As all the SBs are expected to behave in much the same way, 
the Be/Fe ratio, for instance, obtained at the end of any SB lifetime 
should also be approximately the same, and all the individual dilution 
lines overlap to give a general apparent linear growth of Be/Fe as a 
function of Fe/H, as seen from the observational data.

Interestingly enough, this special feature of the SB model can be 
tested observationally.  Indeed, if the distribution of halo stars 
over the metallicity range $10^{-4}$--$10^{-1}Z_{\odot}$ is the result 
of dilution lines, then the same number of stars should be expected at 
all $Z$.  This is in contrast with the expectations of a scenario in 
which Be, B, O and Fe abundances build up continuously and 
monotonically in the Galaxy.  Indeed, in such a model, more stars 
should be expected at low metallicity, because their formation would 
correspond to a time when the star formation rate was higher (more gas 
in the Galaxy).  In fact, the SB model even predicts an increasing 
number of stars at increasing metallicity, because the new generation 
of stars (at the end of the SB lifetime) should be distributed from 
$Z_{\mathrm{ISM}}$ to $Z_{\mathrm{SB}}\sim 10^{-1}Z_{\odot}$, and 
$Z_{\mathrm{ISM}}$ irremediably increases.  Of course, a strong 
observational bias may make this prediction rather hard to test, as it 
is easier to measure very small abundances at higher metallicity.  
However, a more systematic observational work with increased numbers 
of stars (and thus improved statistics) may lead to a definite test of 
the SB model in the future.

Our model also allows one to evaluate the expected dispersion of the 
stars in the data.  Varying the SB parameters, we were able to obtain 
a range of Be/O ratios as the final output of the model, all of them 
laying in the observationally allowed numbers.  We first investigated 
OB associations containing less than 100 SNe (our standard value).  
There are two different ways to do so: i) keeping a SN explosion rate 
of one every $3\,10^{5}$~yr, but shortening the SB lifetime, or ii) 
keeping a SB lifetime of 30~Myr, but lowering the explosion rate and 
thus the SB mechanical power and the power imparted to the EPs.  We 
explored the first possibility down to lifetimes of 10~Myr, and found 
and increase of the Be/O and Be/Fe ratios by about 20\%.  This is 
mostly due to a correspondingly higher reduced
metallicity of the EPs (see Fig.~\ref{zeta}).  As for the second 
possibility, we investigated mechanical powers down to 
$10^{37}\mathrm{erg~s}^{-1}$ (i.e.  one SN every 3~Myr) and found Be/O 
and Be/Fe ratios about 40\% lower, in agreement with the results shown 
in Fig.~\ref{zeta}.

Another cause for the dispersion of the data is expected from the 
overlap of dilution lines corresponding to SBs evolving in an 
interstellar medium of different metallicity. We run the model for an 
ambient metallicity of $10^{-2}Z_{\odot}$, using the corresponding SN 
models of WW95 (models TA), and found Be/O and Be/Fe ratios about 40\% 
higher than for a metallicity of $10^{-4}Z_{\odot}$.  Gathering all 
these results, we finally expect a dispersion of about a factor of 2 
or 3 in the data, corresponding to stars formed from the gas processed 
by SBs created by OB associations with different characteristics.

Note also that, in our calculations, we have used an idealized SB 
model.  However, the results should not be very different for a real 
SB, as far as LiBeB production is concerned.  In particular, the time 
history of the energy release in the OB association is not very 
important, as a given amount of energy imparted to EPs with a given 
composition, always leads to approximately the same integrated LiBeB 
production, independently of the rate at which the energy is injected.  
As for the EP composition, it mainly depends on the amount of material 
evaporated off the shell, as compared with the total mass of the 
ejecta.  But both are directly linked to the number of SNe exploding 
within the SB, and do not depend (or little) on the rate of 
explosions.  To put it in different words, no matter what the exact 
explosion rate is, one should always be able to define a modified time 
coordinate (`renormalized' time), $t^{\prime}$, such that 
$\mathcal{P}^{\prime} \equiv \d E_{SN}/\d t^{\prime}$ is constant and 
the SB evolution laws Eq.~(\ref{RSB})-(\ref{TSB}) approximately hold 
with this new time coordinate.  Our calculations of the LiBeB 
production would thus give results very close to those obtained in the 
genuinely continuous model.

Finally, we argue that apart from the intrinsic dispersion in the data 
discussed above, one should also find in the halo some stars formed 
from a part of the ISM which has not been processed by SBs.  This set 
of stars is expected to show a very different behavior, since it 
results from the activity of individual, isolated SNe, rather than 
collective ones, in an OB association.  Some LiBeB production should 
nevertheless accompany the SN energy release, but it now occurs within 
supernova remnants (SNRs), not superbubbles.  Now the LiBeB production 
within isolated SNRs has been calculated in detail in precedent papers 
(Parizot \& Drury, 1999a,b), and shown to lead to Be/O and Be/Fe 
ratios much lower (by about one order of magnitude) than those 
obtained in the SB model presented here.  These isolated SNe, however, 
and the accompanying LiBeB production process, have no reason not to 
occur, and should be responsible for the general, continuous increase 
of the ISM metallicity.  But the stars formed from this gas should be 
expected to lay on a specific part of the diagram showing, say, the 
Be abundance as a function of Fe/H, namely on a line about one order 
of magnitude lower than the line corresponding to the stars formed 
from the SB processed gas.  This is another prediction of the model 
(unless \emph{all} the SNe occurred in OB associations in the early 
Galaxy, which is not very likely).  Unfortunately, the B and Be 
abundance is obviously very hard to measure in these stars, since it 
is expected to be so low.  Most probably, observations will provide 
upper limits on these abundances.  In this respect, it is very 
interesting to note that only upper limits have been reported for 7 
stars in the sample gathered by Fields \& Olive (1999).  Although 
more observational work and proper statistics would be needed, these 
stars might represent a first piece of evidence in favour of a 
`bimodal' LiBeB production in the Galaxy, that is to say from 
\emph{correlated} and \emph{isolated} SN explosions.

\begin{acknowledgements}
We wish to thank Andrei Bykov and Jean-Paul Meyer for comments 
improving the content as well as the presentation of the paper.  This 
work was supported by the TMR programme of the European Union under 
contract FMRX-CT98-0168.
\end{acknowledgements}

\end{document}